\documentclass[conference]{IEEEtran}
\usepackage{indentfirst}
\usepackage{amsmath}
\usepackage{amsfonts}
\usepackage{dsfont}
\usepackage{cite}
\usepackage{times}
\usepackage[pdftex]{graphicx}
\usepackage{subfigure}
\usepackage{bbm}

\newcommand{\figw}{0.95\columnwidth}
\newcommand{\figww}{0.7\columnwidth}

\title{Content-Based Interference Management for\\ Video Transmission in D2D Communications\\ Underlaying LTE}
\author{Sabur Baidya and Marco Levorato\\
\normalsize $*$ Donald Bren School of Information and Computer Sciences, UC Irvine\\
\normalsize e-mail: \{sbaidya,~levorato\}@uci.edu}
\date{}

\begin{document}

\pagestyle{empty}
\thispagestyle{empty}

\maketitle

\begin{abstract}
A novel interference management approach is proposed for modern communication scenarios, where multiple applications and networks coexist on the same channel resource. The leading principle behind the proposed approach is that the interference level should be adapted to the content being transmitted by the data links to maximize the amount of delivered information. A network setting is considered where Device-to-Device (D2D) communications underlay a Long Term Evolution (LTE) link uploading video content to the network infrastructure. For this scenario, an optimization problem is formulated aiming at the maximization of the D2D link's throughput under a constraint on the Peak Signal-to-Noise-Ratio of the video data stream.
The resulting optimal policy focuses interference on specific packets within the video stream, and significantly increases the throughput achieved by the D2D link compared to an undifferentiated interference strategy. The optimal strategy is applied to a real-world video streaming application to further demonstrate the performance gain.

\end{abstract}

\section{Introduction}

Emerging technological trends such as Urban Internet of Things (IoT) and Smart Cities~\cite{ZorziIOT,bellavista2013convergence,schaffers2011smart} are revolutionizing the network scenario. An increasing
number of diverse applications will coexist on a heterogeneous network infrastructure, thus posing important technical challenges. The multi-scale nature of information acquisition and computation~\cite{bonomi2012fog,edge_mining_2013} that characterizes these applications matches the evolution of the network infrastructure towards multi-scale communications.

Device-to-Device (D2D) communications have been envisioned as an effective solution to support local information exchange without generating additional traffic to the network infrastructure. Herein, we consider a network scenario where a D2D link operates on a Long-Term Evolution (LTE) uplink channel~\cite{boccardi2014five,5350367,6364738,fodor2012design}. 

The main challenge is the control of mutual interference, especially due to the logical separation between the two networks. Recent contributions proposed interference control mechanisms, where the objective is to maintain the Signal-to-Interference-plus-Noise-Ratio (SINR) at the LTE receiver above a predefined threshold~\cite{6364738,janis2009device,7417130}. In this paper, we propose a novel content-based interference strategy, which targets interference on specific packets within the information stream. We contend that this construction is especially suited to the Urban IoT, where data streams from sensors are compressed, and the relevance of the transmitted information might change over time.

Herein, we specifically focus on applications based on video streaming, such as smart camera systems for real-time monitoring.
This choice is motivated by the importance of these systems as a component of smart transportation (\emph{e.g.}, traffic monitoring~\cite{bramberger2004real}) and surveillance systems~\cite{zander2005automated,fleck20063d} in the urban IoT, as well as by the inherent challenges associated with transmitting video streams over wireless~\cite{fitzek2001mpeg}. We consider a scenario where the LTE link is transporting compressed video, and the D2D transmitter is bound by a constraint on the minimum Peak Signal-to-Noise-Ratio (PSNR) measured at the LTE receiver.

The dynamics of the network are modeled as a Markov process~\cite{bert2}, which tracks the transmission process of the LTE link and the transmission decisions of the D2D transmitter. The optimization problem is formulated as a Markov Decision Process (MDP) problem, and solved by means of a specialized Linear Program (LP). The output of the optimization problem is a randomized past-independent policy~\cite{ross_constr}, that is, the transmission action is solely a function of the current state of the system. 

The obtained optimal transmission strategy and achieved performance demonstrate that the D2D link can significantly improve its throughput if interference is targeted to specific packets within the video stream. More specifically, the D2D transmitter should reduce interference generated to the LTE link when reference frames are transmitted. Results obtained by applying the optimal policy to real-world video show that the PSNR significantly improves with respect to that of the reference transmission policy.

The rest of the paper is organized as follows. Section~\ref{sec:net} describes the heterogeneous D2D-LTE network scenario and the video application scenario considered herein. Section~\ref{sec:stocmod} presents the stochastic process used to model the operations of the network. In Section~\ref{sec:perf}, the performance metrics and optimization problem are formulated. Section~\ref{sec:impl} discusses the practical implementation of the optimal policy in real-world networks. Section~\ref{sec:numres} presents numerical results based on the proposed framework. Section~\ref{sec:concl} concludes the paper.

\section{Network and Application Scenario}
\label{sec:net}

\begin{figure}[!t]
	\centering
	\includegraphics[width=\figww]{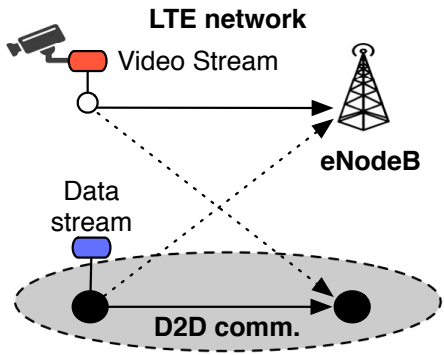}
	\caption{Network configuration considered in the paper. The LTE and D2D links coexist on the same channel resource and mutually interfere.}
        \label{fig:network}
\end{figure}

Fig.~\ref{fig:network} depicts the network considered in this paper, where LTE and D2D links coexist on the same channel resource and mutually interfere. The LTE mobile terminal is uploading a video to the network infrastructure, whereas the D2D device is assumed to be transmitting best-effort traffic to a neighboring mobile device. In the following, we first describe the video compression techniques that are widely used for storage and transmission, and, then, the physical layer model adopted in the analysis and optimization framework.

\subsection{Video Compression and PSNR}

Temporal compression exploits the significant similarities that may interest pictures within the video stream captured at close time instant. Thus, a considerable compression rate can be achieved by transmitting differences with respect to a reference frame or motion vector. This gives the opportunity to encode the video in a small number of key (reference) frames and a larger number of compressed predicted frames following the reference frame. When the decoder receives the predicted frames, it uses the preceding reference frame information to decode each of the predicted frames. As a result, if a reference frame is corrupted, the effect propagates through the entire Group of Pictures (GoP). Alternately, if a predicted frame is damaged, the effect is not so severe compared to the loss of a reference frame. The most popular coding technique is H.264/AVC, which uses a Group of Pictures (GoP) for predictive coding of frames. The GoP starts with intra-coded frame (I-frame) followed by a number of inter coded frames \emph{e.g.}, Predicted frames (P-frames), Bi-directional predicted frames (B-frames). In the following, we refer to P and B-frames as D-frames. For H.264/AVC the GoP can be fixed (constant size) or Adaptive (variable size). Herein, we contend that the different amount of information conveyed by packets associated with I and D-frames should correspond to a different interference level.

\subsection{Physical Layer}

Analogously to most prior literature on D2D underlaying LTE networks, we abstract the physical layer using a per-packet decoding threshold. Let's denote the LTE and D2D transmitter and receiver by the subscript $\ell$ and $d$, respectively.
Due to mutual interference, the Signal to Noise plus Interference Ratio (SINR) at the LTE and D2D receiver are
\begin{equation}
{\rm SINR}_{\ell} = \frac{P_{\ell} c_{\ell\ell} }{\sigma^{2}_{d}+P_{d}  c_{d\ell}} , \hspace{0.3cm} {\rm SINR}_{d} = \frac{P_{d} c_{dd} }{\sigma^{2}_{d}+P_{\ell}  c_{\ell d}} ,
\end{equation}
respectively, where $P_{\ell}$ and $P_{d}$ is the average received power at the LTE and D2D receivers (including path loss), $c_{xy}$ is the fast fading gain from
node $x{\in}\{\ell,d\}$ to receiver $x{\in}\{\ell,d\}$ and $\sigma^{2}_{\rm x}$ is the noise variance at receiver $x$. 

We assume that the coefficients $c_{xy}$ have Rayleigh probability density function, that is,
\begin{equation}
F_{c_{xy}}(c)=\mathit{P}(c_{xy} < c) = 1-e^{-c}, \quad c \geq 0,
\end{equation}
where the power of the fading process is $1$.
Here, we consider a slotted time model, where the fading coefficient is assumed constant within each slot, and fading coefficients in different slots are independent and identically distributed (\emph{i.i.d.}) random variables. In order to provide an intuitive and insightful model, we assume that each slot corresponds to the transmission of one video frame.
Then, the probability that $SINR_{x}$ is above the decoding threshold $\gamma$ and the packet is successfully decoded is
\begin{equation}
\rho_{x}{=} \frac{e^{-\frac{\gamma  \sigma^2_x }{P_x}}}{1+\frac{\gamma  P_y}{P_x}},
\end{equation}
with $x{\neq}y$.
Note that LTE may fragment frames into multiple packets. In this case, our assumption corresponds to invariant fading for the duration of the transmission of the burst of packets associated with a single video frame.

\section{Stochastic Model}
\label{sec:stocmod}

We develop a stochastic model to study the performance of the wireless nodes in the coexistence scenario described earlier, and optimize the transmission strategy of the D2D transmitter. Our objective is to measure the PSNR of the video link and the throughput of the D2D link as a function of the transmission strategy of the D2D transmitter. Clearly, one of the key aspects to capture is the error propagation effect due to the differential encoding used in video compression. In this analytical framework, we assume that the loss of a reference frame causes the loss of all the frames in the GoP. A numerical study of this effect in real-world video streams is provided in Section~\ref{sec:numres}. The definition of models capturing bidirectional dependencies are left to future studies.

\begin{figure}[!t]
	\centering
	\includegraphics[width=\figw]{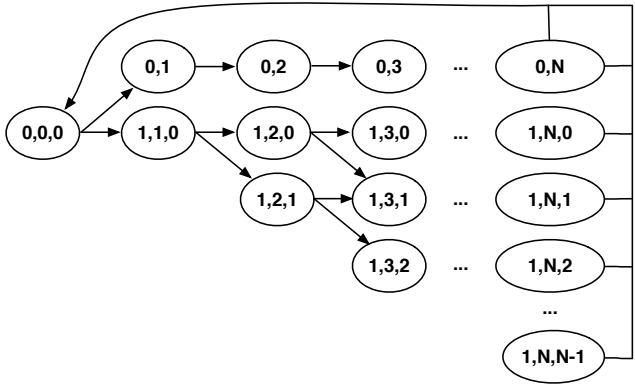}
	\caption{Graphical representation of the Markov chain for a case with fixed GOP length.}
        \label{fig:chain}
\end{figure}

We build the stochastic model to track the number of frames transmitted and delivered within each GoP. We logically divide the temporal evolution of the process into renewal periods where the first state corresponds to the transmission of an I frame, and the subsequent states in each period correspond to the transmission of differential frames. Note that the D2D link is a best effort data stream, where we only track the delivery of individual packets. We, then, define the Markov process ${\bf S}{=}(S(0),S(1),S(2),\ldots)$, where $S(t){\in}\mathcal{S}$ is the state in slot $t$ and $\mathcal{S}$ is a finite state space.
The state of the system is described by the vector $(I_{\rm rx},N_{\rm tx},N_{\rm rx})$. The variable $I_{\rm rx}{\in}\{0,1\}$ is equal to $1$ if the $I$ frame associated with the current GoP was correctly decoded at the video stream destination, and is $0$ otherwise. $N_{\rm tx}$ and $N_{\rm rx}$ are the number of differential frames transmitted and received within the $GoP$, respectively, with $0{\leq}N_{\rm rx}{<}N_{\rm tx}{\leq}N$, where $N$ is the maximum GoP size.
The, possibly variable, size of the GOP is modeled by defining the distribution $\beta$, where $\beta(i)$ is the probability that the GOP terminates after the transmission of $i$ differential frames, with $\beta(N){=}1$. Note that fixed size GOP encoding can be modeled by setting $\beta(i){=}0$ for $i{=}1,\ldots,N-1$. We also define the action variable $U(t){\in}\{0,1\}$, where $0$ and $1$ correspond to idleness and transmission of the cognitive D2D transmission, respectively. Note that different transmission power can be incorporated as decision variables. Herein, we fix the transmission power of the two terminals, and limit the decision of the cognitive terminal to a binary variable. Then, we define $\rho_{x}(u)$ as the failure probability of the packet sent by link $x{\in}\{\ell,d\}$ conditioned on the decision variable $u$. Note that $\rho_{d}(0)$ is trivially equal to $0$, whereas $\rho_{\ell}(0){=}1 - e^{-\frac{\gamma  \sigma^2_{\ell} }{P_{\ell}}}$. When the links mutually interfere, the failure probabilities are
\begin{equation}
\rho_{d}(1){=}1{-} \frac{e^{-\frac{\gamma  \sigma^2_d }{P_d}}}{1+\frac{\gamma  P_{\ell}}{P_d}},~~\rho_{\ell}(1){=}1{-} \frac{e^{-\frac{\gamma  \sigma^2_{\ell} }{P_{\ell}}}}{1+\frac{\gamma  P_d}{P_\ell}}.
\end{equation}
Fig.~\ref{fig:chain} depicts the state space and allowed transitions.

The state $(0,0,0)$ corresponds to the transmission of the I-frame. Conditioned on action $u$, the process then moves either to $(1,1,0)$ or $(0,1,0)$ with probability $1{-}\rho_{\ell}(u)$ and $\rho_{\ell}(u)$, respectively. The former and latter state indicates successful and failed decoding of the I-frame.

From state $(i_{\rm rx},n_{\rm tx},n_{\rm rx})$, $0{<}n_{\rm tx}{<}N$, and conditioned on action $u$, the Markov chain moves to the states
\begin{align}
(0,0,0) &{\rm ~w.p~} \beta(n_{\rm tx}),\\
(i_{\rm rx},n_{\rm tx}{+}1,n_{\rm rx}{+}1)&{\rm ~w.p~} (1{-}\beta(n_{\rm tx}))\rho_{\ell}(u) ,\\
(i_{\rm rx},n_{\rm tx}{+}1,n_{\rm rx})&{\rm ~w.p~} (1{-}\beta(n_{\rm tx}))(1{-}\rho_{\ell}(u)) ,
\end{align}
where $\beta(n_{\rm tx})$ is the probability that the GoP terminates at the transmission of the $n_{\rm tx}$ differential frame, and another I-frame is sent.
From state $(i_{\rm rx},N,n_{\rm rx})$, the process moves to the states $(0,0,0)$ with probability $1$.
since the GoP terminates. Note that in the transition probabilities listed above, the selection mechanism for the control variable is left unspecified. In the next section, we will formulate an optimization problem whose output is the control action distribution defined on the state space of the network.

\section{Performance Metrics and Optimization Problem}
\label{sec:perf}

\subsection{PSNR Analysis}

Our first goal is to derive an evaluation metric, \emph{i.e.},
the PSNR, which is commonly used to measure
video quality. We, then, compute the Mean
Square Error (MSE) of the video sequence averaged over all
frames, and from this value we obtain the PSNR as
\begin{equation}
{\rm PSNR} = 10 \log10({K}_{\rm bps}/{\rm MSE}),
\end{equation}
where ${K}_{\rm bps}$ is $2^W$ and $W$ is the number of bits per pixel.

The technique we use for the evaluation of the PSNR is an
extension of that proposed in~\cite{stuhlmuller2000analysis,badia2010video}. We briefly sketch it and then discuss how it
applies to our evaluations. In~\cite{stuhlmuller2000analysis}, the MSE of the received
video is seen as the sum of multiple uncorrelated distortion
values. Hence, we write $MSE {=} \mathcal{D}_e {+} \mathcal{D}_u$, where
 $\mathcal{D}_e$ is the distortion introduced by the encoder and $ \mathcal{D}_u$ is
the distortion term caused at the decoder by residual errors, \emph{i.e.}, packet error.
Herein, we only focus on $\mathcal{D}_u$ and assume $\mathcal{D}_e$ as constant,
as it does not depend on packet and frame loss. According to the framework proposed in
~\cite{stuhlmuller2000analysis}, which has been validated via extensive simulations, pixel errors caused by not delivered frames can be approximated by Gaussian distributions. Moreover, it is also assumed that further manipulations of the received signal performed by the codec are linear and time-invariant, and thus can be represented through the frequency responses of some filters. If this filtering
is applied iteratively, based on the central limit theorem, one can expect that the impulse response of the filter also becomes
Gaussian after a sufficiently large number of iterations [7]. Thus, we model the term $\mathcal{D}_u$ as a 
Gaussian variable with zero mean and variance $\sigma^2_u$, where the latter linearly depends on the frame error rate.
Following the considerations in~\cite{badia2010video}, this assumption leads to $\mathcal{D}_u{=}C \sigma_u{=} C p_{\rm err} \sigma_e$, where $C$ and $\sigma_e$ are constants and function of the video and encoding/packetization implementation, and  $p_{\rm err}$ is the frame error rate.

\subsection{Performance Metrics}

The PSNR, then, is monotonically increasing in the frame delivery rate, which we use to evaluate the performance of the video. Within the adopted markovian framework, the frame delivery rate is measured as 
\begin{equation}
\mathit{D}_{LTE}(\mu)= \lim_{T{\rightarrow}\infty}\frac{1}{T}\mathit{E}\bigg[ \sum_{t=0}^{\infty} \mathbbm{1}( \Omega(t)) \bigg],
\end{equation}
where $\Omega(t)$ is the event corresponding to a frame delivery in slot $t$. $\mathit{D}_{LTE}(\mu)$ can be rewritten as
\begin{equation}
\mathit{D}_{LTE}(\mu)= \sum_{s{\in}\mathcal{S},u{\in}\{0,1\}} \pi_{\mu}(s,u) \omega(s,u),
\end{equation}
where the function $\omega(s,u)$ counts the frames transmitted in the current GoP, that is,
\begin{equation}
\omega(s,u){=}\beta(n_{\rm tx})*i_{\rm rx} * ( n_{\rm rx} {+} i_{\rm rx}{+} 1{-}\rho_{\ell}(u))
\end{equation}
in state $s{=}(i_{\rm rx},n_{\rm tx}{+}1,n_{\rm rx}{+}1)$. The term $\beta(n_{\rm tx})$ accounts for the fact that $n_{\rm rx}$ is the number of frames received in the buffer only if the frame terminates. The multiplicative term $i_{\rm rx}$ forces to zero the number of frames delivered if the $I$ frame failed.

The normalized throughput of the D2D link, measured in successfully delivered packets per slots, is defined as the long-term average
\begin{equation}
\mathit{T}_{D2D}(\mu)= \lim_{T{\rightarrow}\infty}\frac{1}{T}\mathit{E}\bigg[ \sum_{t=0}^{\infty} \mathbbm{1}( \Phi(t)) \bigg],
\end{equation}
where $\mu$ is the control strategy and $\Phi(t)$ is the event corresponding to the delivery of a packet by the D2D link. The above measure can be rewritten as
\begin{equation}
\mathit{T}_{D2D}(\mu) = \sum_{s{\in}\mathcal{S},u{\in}\{0,1\}} \pi_{\mu}(s,u) \phi(s,u),
\end{equation}
where $\pi_{\mu}(s,u)$ is the joint state-action steady state distribution
\begin{equation}
\pi_{\mu}(s,u) = \lim_{t\rightarrow\infty}\mathit{P}_{\mu}( S(t){=}s,U(t){=}u ),
\end{equation}
and $\phi(s,u){=}1{-}\rho_{d}(u)$.

\subsection{Optimization Problem}

Based on the stochastic model and performance metrics defined earlier, we can now define the optimization problem aimed at maximizing the D2D link throughput under constraints on the PSNR. The optimization problem 
\begin{equation}
\label{optprob}
\mu^* = \arg \max_{\mu}  \mathit{T}_{D2D}(\mu) ~~{\rm s.t.}~~  \mathit{D}_{LTE}(\mu){\geq} \delta,
\end{equation}
admits at least one optimal policy $\mu$ in the class of randomized past independent policies~\cite{ross_constr}. We focus on this class of policies, and define the policy $\mu(u,s){=}\mathit{P}(U(t){=}u|S(t){=}s)$. Then, the problem (\ref{optprob}) 
is mapped to the Linear Program (LP)~\cite{ross_constr,ARQ_icc}
\begin{align}
\label{eq:optprob2}
{\bf z}^* =& \arg\min_{{\bf z}}~~ 1-\sum_{s,\omega} \phi(s,u) \, z(s,u)\\
{s.t.}~~& \sum_{s,\omega} \psi(s,u) \, z(s,u) {\leq} \delta \nonumber\\
& \sum_{s,u} z(s,u) p(s^{\prime}|s,u) = \sum_{\omega\in\mathcal{A}} z(s^{\prime},u),\nonumber\\
& \sum_{s,u} z(s,u) = 1,\nonumber\\
& z(s,u) \geq 0, ~\forall s,\omega,\nonumber
\end{align}
where for the sake of readability, we denote $\sum_{s\in\mathcal{S}}\sum_{u\in\{0,1\}}$
with $\sum_{s,u}$ and ${\bf z}$ as the vector $\{z(s,u)\}_{\forall s,u}$, and $\mathit{p}(s^{\prime}|s,u)$ is the transition probability to $s^{\prime}$ conditioned on the state $s$ and action $u$. In these formulation, the optimization variable $z(s,u)$ is the the joint steady-state probability 
of the state-action pair $(s,u)$, that is, $z(s,u)=\Pr(S(t){=}s,U(t){=}u)$. The optimal policy $\mu^*$, then, is
\begin{equation}
\mu^*(s,u)= \frac{z(s,u)}{\sum_{u^{\prime}\in\mathcal{A}}z(s,u^{\prime})} ~~\forall s,u.
\end{equation} 

\section{Implementation of the Optimal Policy}
\label{sec:impl}

The high level of abstraction of the proposed model is used to derive guidelines for the definition of adaptive transmission protocols in heterogeneous application and network scenarios. Clearly, the application of such policies in a real-world scenario necessitates advanced capabilities of the terminals and network devices, and should account for some important networking aspects.

Firstly, we observe that in practical networks, the interference strategy needs to be defined at the LTE Medium Access Control (MAC) level, where video frames are fragmented into several LTE packets. These packets are, then, transmitted over the channel. Heuristics based on the optimal policy can be directly applied to this case, where the transmission probability is a function of whether the packets being transmitted belong to an I-frame or D-frame, as well as whether or not the I-frame in the GOP was significantly damaged.

A critical point in the practical implementation of this class of policies is the availability of ``state'' information at the D2D transmitter. Importantly, different from prior work, where SINR control requires precise channel knowledge and power control loops, our policies are based on statistical knowledge of the channel, that is, the failure probabilities. 

However, even if coarse heuristics are used, the D2D transmitter needs to tune the transmission probability on the type of frame being transmitted. Since LTE packets are encrypted, the D2D transmitter cannot acquire this information by overhearing the transmitted LTE packets, and necessitates the assistance of the LTE eNodeB. 
In the framework proposed herein, the optimal action distribution is defined on the specific state within the state space of the Markov process. Thus, the implementation of such policy may require an excessive amount of signaling. However, heuristic policies can be devised to reduce the information needed at the D2D transmitter, by defining the action distribution solely on the class of the frame being transmitted (that is, I-frames and D-frames). Unless specific markers are added to the packets, the eNodeB, or control units connected to the eNodeB, should decode the packets to identify frame end/begin and the type of frame. Thus, the proposed policies need  \emph{deep packet inspection}~\cite{smith2008deflating}, which has been receiving a considerable amount of attention by the research community as an effective technique to provide novel network services.

\section{Numerical Results}
\label{sec:numres}

We now present numerical results assessing the performance of the proposed interference scheme. We compare the optimal policy to a baseline case where the transmission probability of the D2D transmitter is equal to $p_{\rm tx}$ in all the states.
In this case, for fixed GOP size, the delivery rate of frames is
\begin{align}
\mathit{D}_{LTE}(p_{\rm tx})&{=}\left[\sum _{i=0}^{N} \frac{N!}{i! (N-i)!} (1-\rho_{\ell}(p_{\rm tx}) )^{i} \rho_{\ell}(p_{\rm tx})^{N-i}\right]\nonumber\\
&{=}\frac{(1-\rho_{\ell}(p_{\rm tx}) ) (N (1-\rho_{\ell}(p_{\rm tx}) )+1)}{N+1},
\end{align}
where 
\begin{equation}
\rho_{\ell}(p_{\rm tx})= \rho_{\ell}(1) p_{\rm tx} + \rho_{\ell}(0) (1{-}p_{\rm tx}).
\end{equation}
The throughput of the D2D link is trivially equal to 
\begin{equation}
\mathit{T}_{D2D}(p_{\rm tx}) = p_{\rm tx}(1-\rho_{d}(1)).
\end{equation}

\begin{figure}[t]
	\centering
	\includegraphics[width=\figw]{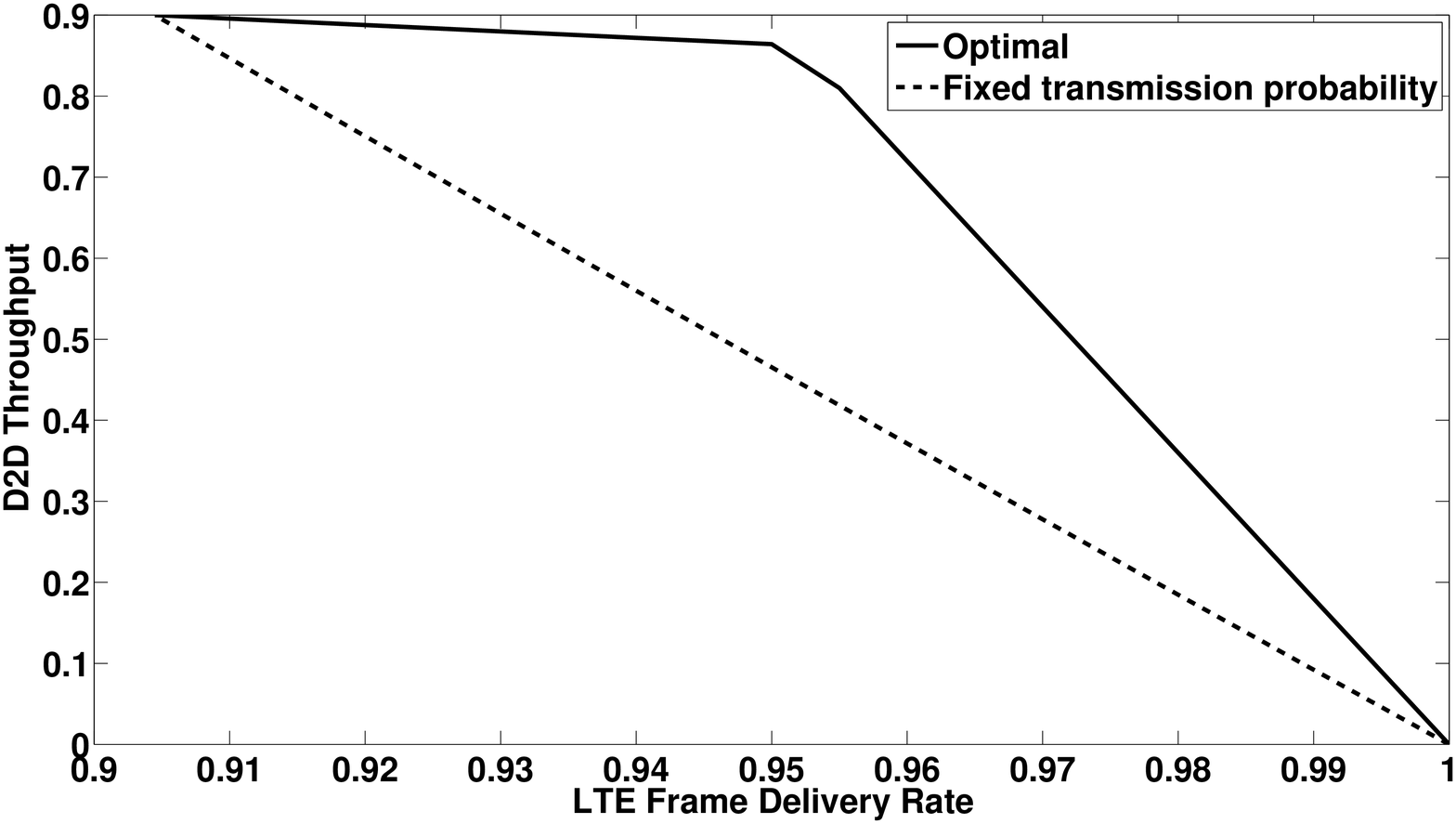}
	\caption{Throughput of the D2D link as a function of the LTE frame delivery rate for the optimal (solid line) and the fixed transmission probability (dashed line) policy.}
        \label{fig:perf}
\end{figure}

In the following plots, we assume a fixed GOP size equal to $24$ frames, including the I-Frame. The error probabilities are set to $\rho_{\ell}(0){=}0.01$, and $\rho_{\ell}(1){=}\rho_{d}(1){=}0.1$. Fig.~\ref{fig:perf} shows the throughput of the D2D link as a function of the LTE frame delivery rate for the optimal (solid line) and the fixed transmission probability (dashed line) policy.
In the fixed transmission probability case, the throughput of the D2D link decreases linearly as the delivery rate of the LTE frames increases. The optimal policy significantly increases the throughput of the D2D link until the constraint on the LTE frame delivery 
becomes too tight, and the D2D transmitter is forced idle.

\begin{figure}[!t]
	\centering
	\includegraphics[width=\figw]{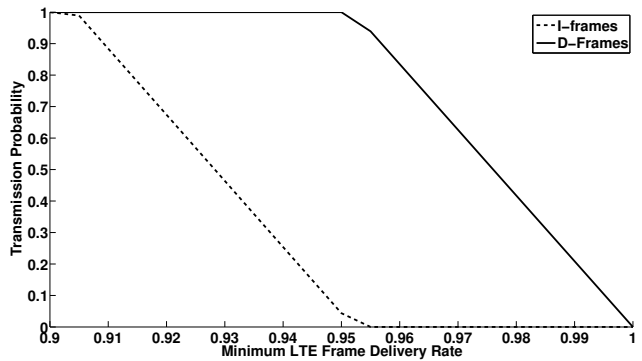}
	\caption{Transmission strategy of the D2D link as a function of the constraint on the minimum LTE frame delivery rate. The dashed and dotted lines correspond to transmission probabilities when an I-Frame and a D-frame are transmitted.}
        \label{fig:pol}
\end{figure}

Fig.~\ref{fig:pol} depicts the transmission strategy of the D2D link as a function of the constraint on the minimum LTE frame delivery rate. The dashed and dotted lines correspond to transmission probabilities when an I-Frame and a D-frame are transmitted. In the latter case, only frames where the I-frame in the same GOP was delivered are considered. In fact, the optimal policy trivially prescribes transmission with probability $1$ in states corresponding to D-frames whose I-frame was lost. When the minimum LTE delivery rate is equal to or larger than the maximum (approximately $0.98$), the D2D transmitter is idle in the state corresponding to I-frame transmission. As the constraint decreases, the D2D transmitter increases the transmission probability in states corresponding to D-frames transmission, whereas the policy prescribes idleness in states corresponding to I-Frame transmission. As the transmission probability in D-frames reaches $1$, the D2D transmitter begin to increase the transmission probability in the state corresponding to I-frame transmission. 
The optimal policy, then, avoids interference to the I-frames to minimize damage to the video until the minimum frame delivery rate is large enough to tolerate the cascade effect triggered by I-frame loss events.

\begin{figure}[!t]
	\centering
	\includegraphics[width=\figw,natwidth=610,natheight=642]{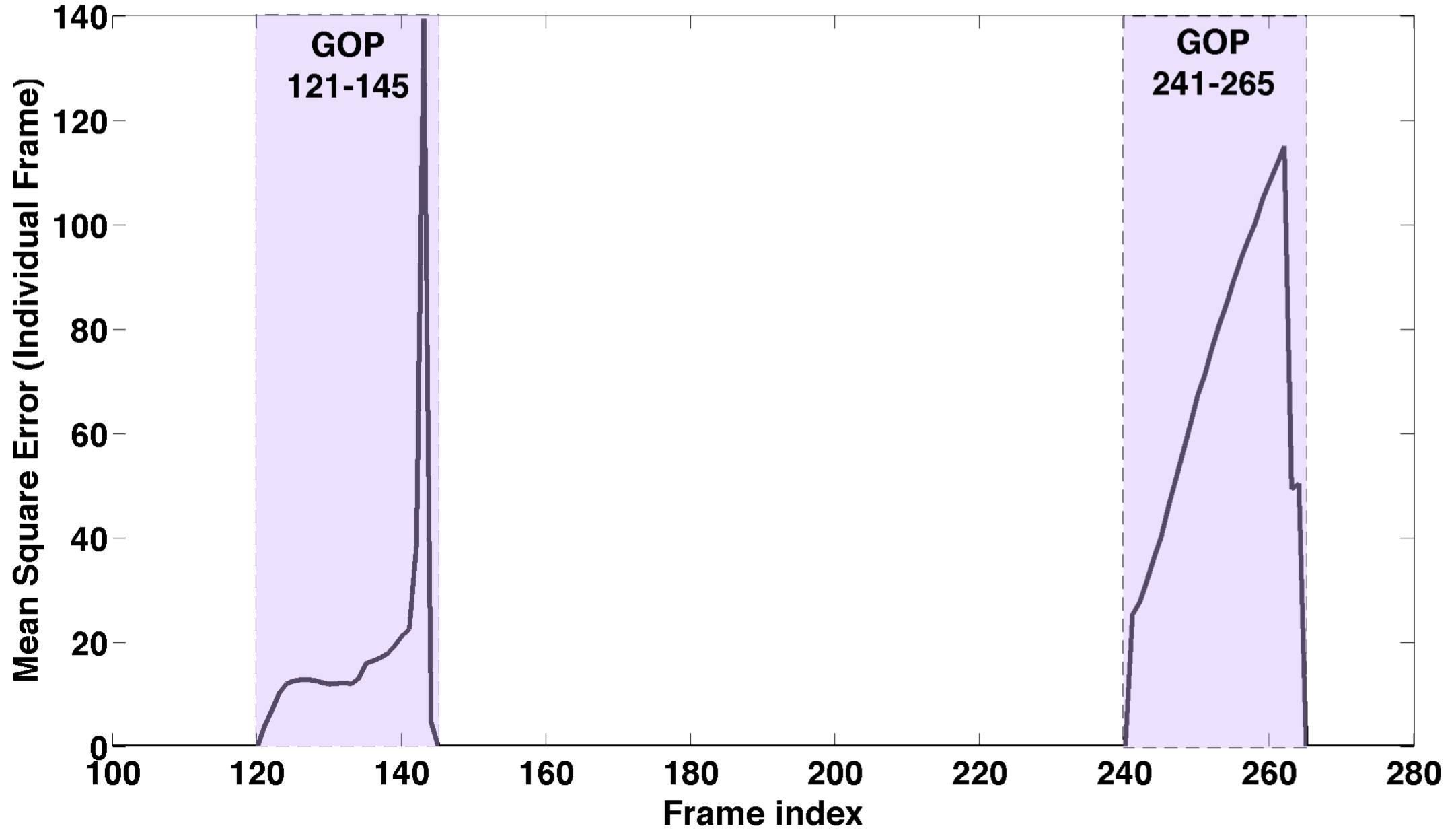}
	\caption{MSE as a function of the frame index. The I-frames with indexes $121$ and $241$ are lost, and the error propagates through the entire GOP ($24$ frames).}
        \label{fig:mse}
\end{figure}

The framework presented in this paper is based on the assumption that the loss of an I-frame impacts the entire GOP. Fig.~\ref{fig:mse} illustrates this effect in a real-world video. The video has fixed GOP length equal to $24$ frames, and size $640{\times}360$ pixels. The I-frames with indexes $121$ and $241$ are lost, and the individual-frame MSE in the frame range $100$ to $280$ is shown. The propagation of the error generated by the loss of the I-frames within their respective GOPs is apparent, thus motivating our model choice.

\begin{figure}[!t]
	\centering
	\includegraphics[width=\figw]{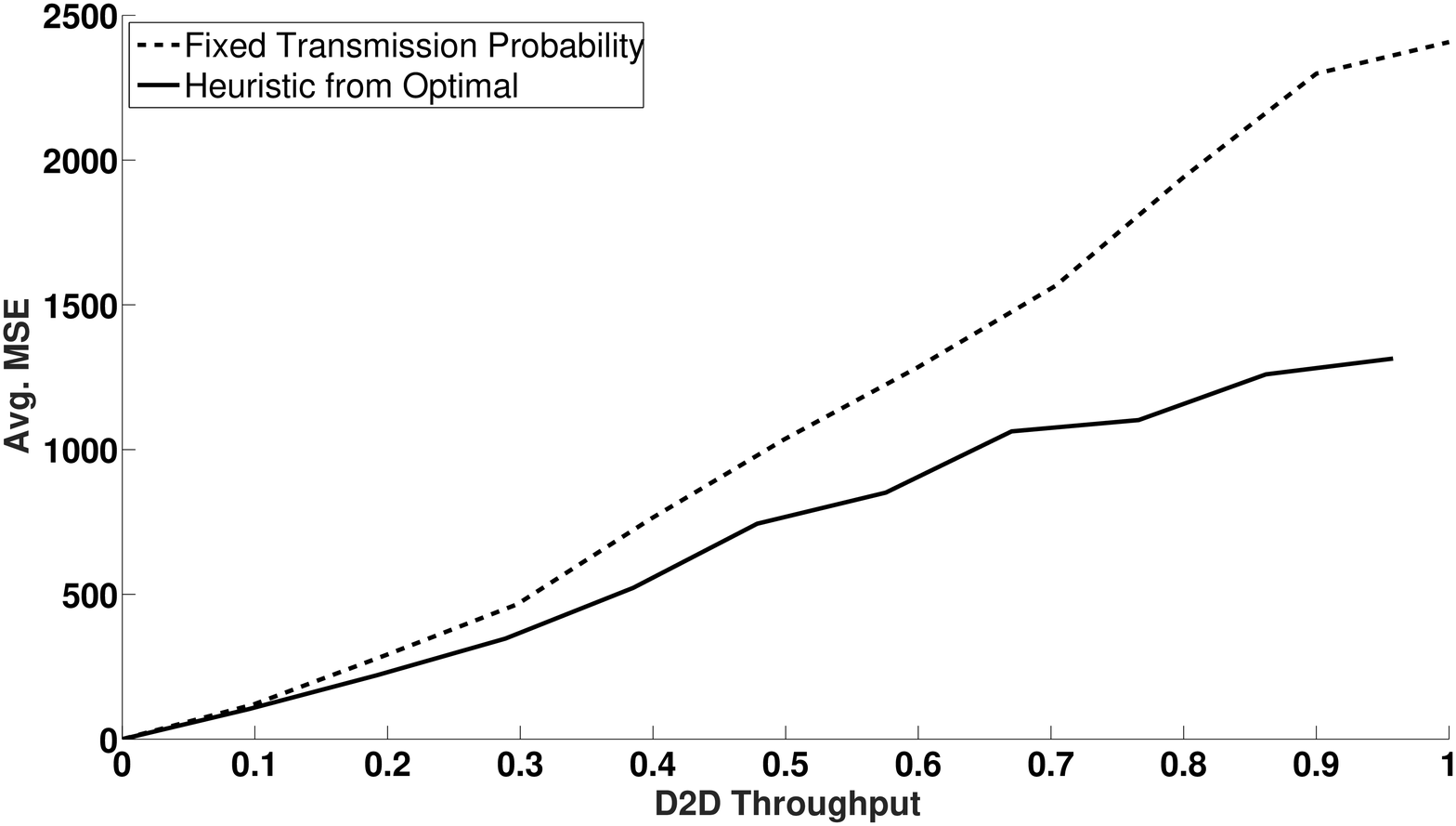}
	\caption{Average MSE as a function of the throughput achieved by the D2D link.}
        \label{fig:mse2}
\end{figure}

Finally, in Fig.~\ref{fig:mse2} we show the MSE as a function of the D2D throughput for the video in the previous picture. The shown results are obtained using the fixed transmission probability policy, and a heuristic extracted from the optimal policy, where the transmission probability is set to $0$ when an I-frame is transmitted, and is constant otherwise. The points in the plot correspond to different values of the transmission probabilities. Note that this policy is simpler to implement in real-world networks, as it requires much less coordination compared to a policy defined on the full state space of the Markov process modeling the dynamics of the network. In the plot, the failure rates were set to $\rho_{\ell}(0){=}0$, $\rho_{\ell}(1)$ and $\rho_{d}(1){=}0$ to increase the impact of D2D transmission on the LTE link and improve convergence rate of computationally demanding simulations. It can be observed that the heuristic policy significantly reduces the MSE, especially in the high throughput region. We remark that the parameters chosen in these simulations result in strong interference to the LTE receiver when the D2D transmitter is active.

\section{Conclusions}
\label{sec:concl}

A novel content-based interference control approach was proposed for scenarios where multiple applications and networks coexist on the same bandwidth. For a network scenario where a D2D link underlays a LTE communication channel where a mobile terminal is uploading a video to the infrastructure. The operations of the network are modeled as a Markov process designed to track the video frames transmitted and delivered by the LTE transmitter. The throughput optimal transmission, and, thus, interference, strategy for the D2D link was obtained by formulating a Markov Decision Process. Numerical results show that, for a given LTE frame delivery rate, the optimal policy significantly increases the throughput of the D2D link, thus facilitating the coexistence of multiple applications on the same channel resource.

\bibliographystyle{IEEEtran}
\bibliography{CNSbib,AF14,cognet15}

\end{document}